\documentclass[aps,preprint,nofootinbib]{revtex4}%
\usepackage{amsfonts}
\usepackage{amsmath}
\usepackage{amssymb}
\usepackage{graphicx}%
\providecommand{\U}[1]{\protect\rule{.1in}{.1in}}

\begin{document}
\preprint{ }
\title[Short title for running header]{Comment on \textquotedblleft Modified F(R) Ho\v{r}ava-Lifshitz gravity: a way
to accelerating FRW cosmology\textquotedblright\ by M. Chaichian, S. Nojiri,
S. D. Odintsov, M. Oksanen, A. Tureanu }
\author{N. Kiriushcheva}
\email{nkiriush@uwo.ca}
\author{P. G. Komorowski}
\email{pkomoro@uwo.ca}
\author{S. V. Kuzmin}
\email{skuzmin@uwo.ca}
\affiliation{The Department of Applied Mathematics, The University of Western Ontario,
London, Ontario, N6A 5B7, Canada}

\begin{abstract}
The partial Hamiltonian analysis of the Ho\v{r}ava-type action presented in
the paper by M. Chaichian, S. Nojiri, S. D. Odintsov, M. Oksanen, A. Tureanu
(\textit{Class. Quant. Grav. 27 (2011) 185021}) is incorrect; for the authors'
choice of variables, a covariant shift, instead of a contravariant shift which
is the one usually used in General Relativity (GR) in ADM variables, the true
algebra of constraints differs from what they presented. The algebra of
constraints for their choice of variables is explicitly given for GR and
compared with the standard algebra.

\end{abstract}
\volumeyear{year}
\volumenumber{number}
\issuenumber{number}
\eid{identifier}
\maketitle

\section{Introduction}

Ho\v{r}ava-type models have become a very popular subject and it is the
custom, as in the original paper \cite{HoravaJHEP2009}, to present some
elements of the Hamiltonian analysis; in this respect paper \cite{MasudCGQ} is
no exception (see section 3). In Ho\v{r}ava-type models, the full
diffeomorphism of the Einstein theory (and Einstein's main idea: general
covariance) is abandoned. This deformation of General Relativity is a topic of
continuing discussion, with disparate opinions.

The Hamiltonian formulation of singular systems can be obtained through a
general procedure that can be applied to covariant and non-covariant systems.
When considering Ho\v{r}ava-type models, the residual GR properties and the
rules of Dirac's procedure should not be abandoned, e.g. the dependence of
Dirac's procedure on the choice of independent variables, and the fact that
raising and lowering indices in GR are performed by fields (of course, for
Ho\v{r}ava-type models only spatial indices are to be raised and lowered).

In paper \cite{MasudCGQ} the Hamiltonian formulation of the Ho\v{r}ava-type
model is performed using ADM variables, but in a form that is non-standard for
GR ($N_{k}$ is taken as an independent variable, in contrast to the standard
choice of $N^{k}$). This departure from convention must affect the Hamiltonian
formulation and the corresponding algebra of constraints. Calculations of some
Poisson Brackets are provided to illustrate that the result does not simply
mimic the standard one. The complete algebra of secondary first-class
constraints of GR in ADM variables for the standard choice ($N^{k}$) is also
compared with the one calculated for the choice of variables used in
\cite{MasudCGQ}.

\section{Re-examination of Hamiltonian analysis of the model of
\cite{MasudCGQ}}

The Hamiltonian analysis (partial) of the modified Ho\v{r}ava-type model (see
(C27))\footnote{We use (C\#\#) to refer to equations of \cite{MasudCGQ}.} was
performed in \cite{MasudCGQ}. After the elimination of the pair of
second-class constraints and the corresponding pair of phase-space variables
($A,\pi_{A}$), the total Hamiltonian (C35)\footnote{Instead of using
undetermined Lagrange multipliers in front of the primary constraints in
(\ref{eqn1}), we write the undetermined velocities as they appear in the
Legendre transformation. In the case under consideration this difference is
not important, but for some formulations it is crucial for the restoration of
the gauge transformations of all fields of an action.} was obtained:%
\begin{equation}
H_{T}=\dot{N}\pi+\int d\mathbf{x}\dot{N}_{i}\pi^{i}+N\int d\mathbf{x}%
H_{0}+\int d\mathbf{x}N_{i}H^{i}. \label{eqn1}%
\end{equation}

Note that the independent variables (and conjugate momenta) of this
formulation are $N(\pi)$, $N_{i}(\pi^{i})$, $g_{pq}(\pi^{pq})$, $A\left(
\pi_{A}\right)  $ and $B\left(  \pi_{B}\right)  $. For this non-standard
choice of variables (compared with GR in ADM variables) the fundamental
Poisson Brackets (PBs) are (see (C31) for the full set)%

\begin{equation}
\left\{  g_{ij}\left(  \mathbf{x}\right)  ,\pi^{kl}\left(  \mathbf{y}\right)
\right\}  =\frac{1}{2}\left(  \delta_{i}^{k}\delta_{j}^{l}+\delta_{i}%
^{l}\delta_{j}^{k}\right)  \delta\left(  \mathbf{x}-\mathbf{y}\right)  ,\text{
\ \ \ \ \ \ }\left\{  N_{i}\left(  \mathbf{x}\right)  ,\pi^{j}\left(
\mathbf{y}\right)  \right\}  =\delta_{i}^{j}\delta\left(  \mathbf{x}%
-\mathbf{y}\right)  . \label{eqn2}%
\end{equation}

In the projectable case, $N=N\left(  t\right)  $, the closure of the
constraint algebra is assumed to be of the form (see (C43))%

\begin{equation}
\left\{  \Phi_{0},\Phi_{0}\right\}  =0,\text{ \ }\left\{  \Phi_{S}\left(
\xi_{i}\right)  ,\Phi_{0}\right\}  =0,\text{ \ }\left\{  \Phi_{S}\left(
\xi_{i}\right)  ,\Phi_{S}\left(  \eta_{j}\right)  \right\}  =\Phi_{S}\left(
\xi^{j}\partial_{j}\eta_{i}-\eta^{j}\partial_{j}\xi_{i}\right)  \approx0,
\label{eqn3}%
\end{equation}
where%

\begin{equation}
\Phi_{0}=\int d\mathbf{x}H_{0}~,\text{ \ \ \ \ \ \ }\Phi_{S}\left(  \xi
_{i}\right)  =\int d\mathbf{x}\xi_{i}H^{i}. \label{eqn4}%
\end{equation}

Equations (\ref{eqn3}) are similar to those presented by Ho\v{r}ava
\cite{HoravaJHEP2009}, which are\ based on the results known for GR in ADM
variables with the additional projectability condition, $N=N\left(  t\right)
$; but in the original Ho\v{r}ava paper \cite{HoravaJHEP2009} the constraint
algebra was written for the standard form of shift variable, $N^{k}$, and its
conjugate momentum, $\pi_{k}$ (the primary constraint), that leads to the
secondary constraint $H_{k}$, which is related to $H^{i}$ of \cite{MasudCGQ} by%

\begin{equation}
H_{k}=g_{ik}H^{i}. \label{eqn5}%
\end{equation}

Relation (\ref{eqn5}) differs from that for an ordinary field theory where the
indices are raised and lowered by the Minkowski tensor without affecting the
calculation of the PBs. Further, $\xi_{i}$ and $\eta_{i}$ in (\ref{eqn3}%
)-(\ref{eqn4}) are test functions that appear in smeared or global
constraints. Based on the distributional properties of delta functions, they
merely allow a different formal presentation of the PBs of two constraints
(e.g. see section 3 of \cite{EPJC}). The test functions are assumed to have a
zero PB with all of the variables present in the constraints, but this
property would be lost if one were to use $g_{ik}\xi^{k}$ instead of $\xi_{i}%
$. To state this fact more explicitly, when shift is treated as a test
function in the calculations, the result depends on which variables are
defined as independent (with the fundamental PBs (\ref{eqn2})). For example,
for the fundamental PB of (\ref{eqn2}): $\left\{  N_{i}\left(  \mathbf{x}%
\right)  ,\pi^{kl}\left(  \mathbf{y}\right)  \right\}  =0$. But for the
dependent variable $N^{p}$, one obtains%

\begin{equation}
\left\{  N^{p}\left(  \mathbf{x}\right)  ,\pi^{kl}\left(  \mathbf{y}\right)
\right\}  =N_{q}\left\{  g^{pq}\left(  \mathbf{x}\right)  ,\pi^{kl}\left(
\mathbf{y}\right)  \right\}  =-\frac{1}{2}\left(  N^{l}g^{pk}+N^{k}%
g^{pl}\right)  \delta\left(  \mathbf{x}-\mathbf{y}\right)  \neq0. \label{eqn6}%
\end{equation}

Even without a change of independent variables, if some constraints (e.g.
$H_{k}$) are linear combinations of others (e.g. $g_{ik}H^{i}$), with
field-dependent coefficients (fields that are canonically conjugate to
variables present in constraints), then the constraint algebra cannot be the
same (for a more detailed example see section 4 of \cite{EPJC}). Such a change
does not affect the closure of the algebra, but the form of closure must be
different. Based on the general arguments for the choice of parametrisation in
\cite{MasudCGQ}, which have just been described, the algebra must differ from
what is asserted.

To illustrate this general point we perform simple calculations for the model
of \cite{MasudCGQ}. The total Hamiltonian (\ref{eqn1}) has the following
so-called Hamiltonian and momentum constraints (see (C34)):%
\begin{equation}
H_{0}=\frac{1}{\sqrt{g}}\left[  \frac{1}{B}\left(  g_{ik}g_{jl}\pi^{ij}%
\pi^{kl}-\frac{1}{3}g_{ij}g_{kl}\pi^{ij}\pi^{kl}\right)  -\frac{1}{3\mu}%
g_{pq}\pi^{pq}\pi_{B}-\frac{1-3\lambda}{12\mu^{2}}B\left(  \pi_{B}\right)
^{2}\right]  \label{eqn7}%
\end{equation}

\[
+\sqrt{g}\left[  B\left(  E^{ij}G_{ijkl}E^{kl}+A\right)  -F\left(  A\right)
+2\mu g^{ij}\nabla_{i}\nabla_{j}B\right]  ,
\]

\begin{equation}
H^{i}=-2\partial_{j}\pi^{ij}-g^{ij}\left(  2\partial_{k}g_{jl}-\partial
_{j}g_{kl}\right)  \pi^{kl}+g^{ij}\partial_{j}B\pi_{B}~, \label{eqn8}%
\end{equation}
where the elimination of the second-class constraints\footnote{To simplify the
notation in this comment we omit the superscript $^{\left(  3\right)  }$, i.e.
$g_{km}^{\left(  3\right)  }=g_{km}$, $\sqrt{g^{\left(  3\right)  }}=\sqrt{g}$
.} gives $A=\tilde{A}\left(  B\right)  $. Consider the PB, $\left\{
H^{i}\left(  \mathbf{x}\right)  ,\int d\mathbf{y}H_{0}\left(  \mathbf{y}%
\right)  \right\}  $, that is claimed to be zero in \cite{MasudCGQ} (where the
projectable case is discussed); the PB for one particular term of $H_{0}$
(second in (\ref{eqn7})), i.e.%

\begin{equation}
\left\{  H^{i}\left(  \mathbf{x}\right)  ,\int d\mathbf{y}\left(  -\frac
{1}{3\mu}\right)  \left(  \frac{1}{\sqrt{g}}g_{pq}\pi^{pq}\pi_{B}\right)
\left(  \mathbf{y}\right)  \right\}  , \label{eqn9}%
\end{equation}
produces non-zero contributions, which can be immediately seen for the PB of
the last term of the momentum constraint (\ref{eqn8}):%

\begin{equation}
-\frac{1}{3\mu}\left(  \partial_{j}B\pi_{B}\right)  \left(  \mathbf{x}\right)
\int d\mathbf{y}\left[  \left\{  g^{ij}\left(  \mathbf{x}\right)  ,\pi
^{pq}\left(  \mathbf{y}\right)  \right\}  \left(  \pi_{B}\frac{1}{\sqrt{g}%
}g_{pq}\right)  \left(  \mathbf{y}\right)  \right]  =\frac{1}{3\mu}\frac
{1}{\sqrt{g}}g^{ij}\partial_{j}B\left(  \pi_{B}\right)  ^{2}\left(
\mathbf{x}\right)  . \label{eqn10}%
\end{equation}

Note: all other contributions of (\ref{eqn9}) are linear in momentum $\pi_{B}$
and cannot compensate (\ref{eqn10}), which is proportional to $\left(  \pi
_{B}\right)  ^{2}$. There is only one term in the Hamiltonian constraint
(third term of (\ref{eqn7})) that might contribute a $\left(  \pi_{B}\right)
^{2}$\ term, but it enters (\ref{eqn7}) with a different numerical
coefficient. If contribution (\ref{eqn10}) could be compensated, it would
place a restriction on the coefficients (i.e. a particular relation between
$\mu$ and $\lambda$). Yet, the PB of the momentum constraint with the third
term of (\ref{eqn7}) is%

\begin{equation}
\left\{  H^{i}\left(  \mathbf{x}\right)  ,\int d\mathbf{y}\left(
-\frac{1-3\lambda}{12\mu^{2}}\right)  \left(  \frac{1}{\sqrt{g}}B\pi_{B}%
^{2}\right)  \left(  \mathbf{y}\right)  \right\}  =0, \label{eqn11}%
\end{equation}
and compensation of (\ref{eqn10}) is impossible, even with a restriction on
the parameters $\mu$ and $\lambda$; therefore, the claim made by the authors
that PB (\ref{eqn3}) is zero for their choice of variables ($N_{i}$ and the
corresponding momentum\ constraint $H^{i}$ (\ref{eqn8})) is incorrect.

The structure of contribution (\ref{eqn10}) and the zero value of PB
(\ref{eqn11}) preclude the result that (\ref{eqn9}) is proportional to the
Hamiltonian constraint. To have closure on the secondary constraint, only a
proportionality to the momentum constraint can be expected. Indeed, the
complete calculation of (\ref{eqn9}) confirms such an expectation, i.e.%

\begin{equation}
\left\{  H^{i}\left(  \mathbf{x}\right)  ,\int d\mathbf{y}\left(  -\frac
{1}{3\mu}\right)  \left(  \frac{1}{\sqrt{g}}g_{pq}\pi^{pq}\pi_{B}\right)
\left(  \mathbf{y}\right)  \right\}  =\frac{1}{3\mu}\frac{\pi_{B}}{\sqrt{g}%
}H^{i}\left(  \mathbf{x}\right)  . \label{eqn12}%
\end{equation}

Instead of calculating a complete algebra of constraints for model
\cite{MasudCGQ}, we show how the authors' choice of variables affects the
algebra of constraints for the Hamiltonian of GR in the ADM variables (to the
best of our knowledge, such an algebra was not reported). For the standard
choice of ADM variables, the total Hamiltonian is%

\[
H_{T}=\int d\mathbf{x}\left(  \dot{N}\pi+\dot{N}^{i}\pi_{i}+NH_{0}+N^{i}%
H_{i}\right)  ,
\]
and its algebra is well-known (e.g. see Castellani's paper \cite{Castellani}), i.e.%

\begin{equation}
\left\{  H_{0},\int d\mathbf{y}NH_{0}\right\}  =N_{,r}g^{rm}H_{m}+\left(
Ng^{rm}H_{m}\right)  _{,r}~, \label{eqn14}%
\end{equation}

\begin{equation}
\left\{  H_{0},\int d\mathbf{y}N^{k}H_{k}\right\}  =\left(  N^{k}H_{0}\right)
_{,k}~, \label{eqn15}%
\end{equation}

\begin{equation}
\left\{  H_{k},\int d\mathbf{y}NH_{0}\right\}  =N_{,k}H_{0}~, \label{eqn16}%
\end{equation}

\begin{equation}
\left\{  H_{k},\int d\mathbf{y}N^{m}H_{m}\right\}  =N_{,k}^{m}H_{m}+\left(
N^{m}H_{k}\right)  _{,m}~. \label{eqn17}%
\end{equation}

We choose to work with PBs of this form, (\ref{eqn14})-(\ref{eqn17}) (without
test functions), because it is\ the form used in the demonstration of the
closure of Dirac's procedure on the secondary constraints; further, this form
is needed for the construction of the generator to restore the gauge
transformations of \textit{all} fields, irrespective of how they were named
(dynamical, non-dynamical, multipliers, etc.). The gauge transformations that
follow from the Hamiltonian formulation should be the same as the Lagrangian
transformations found for \textit{all} fields.

The total Hamiltonian for the choice of shift in \cite{MasudCGQ} becomes%
\begin{equation}
H_{T}=\int d\mathbf{x}\left(  \dot{N}\pi+\dot{N}_{i}\pi^{i}+NH_{0}+N_{i}%
H^{i}\right)  , \label{eqn18}%
\end{equation}
and although the constraint algebra is also closed, it is more complicated
than it would be for the standard choice:%
\begin{equation}
\left\{  H_{0},\int d\mathbf{y}NH_{0}\right\}  =N_{,k}H^{k}+\left(
NH^{k}\right)  _{,k}, \label{eqn19}%
\end{equation}

\begin{equation}
\left\{  H_{0},\int d\mathbf{y}N_{k}H^{k}\right\}  =\left(  N_{k}g^{ki}%
H_{0}\right)  _{,i}+2\frac{1}{\sqrt{g}}G_{pqab}\pi^{pq}g^{ka}N_{k}H^{b},
\label{eqn20}%
\end{equation}

\begin{equation}
\left\{  H^{k},\int d\mathbf{y}NH_{0}\right\}  =g^{ki}N_{,i}H_{0}-2N\frac
{1}{\sqrt{g}}g^{kp}G_{abpq}\pi^{ab}H^{q}, \label{eqn21}%
\end{equation}

\begin{equation}
\left\{  H^{k},\int d\mathbf{y}N_{m}H^{m}\right\}  =-\partial_{i}\left(
N_{m}g^{km}H^{i}\right)  -\mathbf{\partial}_{i}N_{m}g^{km}H^{i}+N_{m}%
g^{kn}g^{mj}\left(  \partial_{n}g_{ji}-\partial_{j}g_{ni}\right)  H^{i}.
\label{eqn22}%
\end{equation}

Note: if for the standard choice of variables only (\ref{eqn14}) has a
field-dependent structure \textquotedblleft constant\textquotedblright, then
for the choice in \cite{MasudCGQ} the situation is the opposite; all PBs
(\ref{eqn20})-(\ref{eqn22}), with the exception of (\ref{eqn19}), have
field-dependent structure \textquotedblleft constants\textquotedblright,\ and
they are very complicated. Imposing the projectability condition leads to the
elimination of terms with spatial derivatives of lapse, and of all total
spatial derivatives in (\ref{eqn14})-(\ref{eqn15}) and (\ref{eqn19}%
)-(\ref{eqn20}). For the standard choice of variables one obtains the
Ho\v{r}ava algebra \cite{HoravaJHEP2009}, but for the non-standard choice of
\cite{MasudCGQ}, the algebra\textbf{ }is different. One can observe that the
form of contribution (\ref{eqn12}) is consistent with the second term of
(\ref{eqn21}).

\section{Conclusion}

The Hamiltonian formulation of singular systems is sensitive to the choice of
field parametrisation \cite{Myths} that, in particular, affects the PBs of
constraints with the total Hamiltonian and among themselves. Moreover,
according to the Dirac conjecture, \textit{all} first-class constraints
generate gauge transformations \cite{Diracbook} and a knowledge of their PB
algebra is important for the construction of the gauge generator (e.g. see
\cite{Castellani}). \ 

In \cite{MasudCGQ} (also see papers by the same authors \cite{MasudPRD,
MasudPLB, COTEPJC}, and by others, e.g. \cite{Horava2010}) the non-standard
choice of shift function was made: instead of $N^{k}$, $N_{k}$ was chosen as
the independent variable. Unlike ordinary field theories, even the choice of
covariant or contravariant fields (because another \textit{field,} the metric
$g_{km}$, is used to raise or lower indices) drastically affects the results
of PBs of constraints with the total Hamiltonian and among themselves. The
assumption made in \cite{MasudCGQ} that the algebra of constraints is the same
for both choices of shift function, is incorrect. For example, (\ref{eqn10})
is obviously non-zero, contrary to the claim of \cite{MasudCGQ}. In the case
of GR written in ADM variables, the well known algebra of constraints
(\ref{eqn14})-(\ref{eqn17}), which is always written for the standard choice
of shift ($N^{k}$) function, becomes much more complicated for the choice of
$N_{k}$ in \cite{MasudCGQ}, see (\ref{eqn19})-(\ref{eqn22}).

\bigskip
\end{document}